\documentclass[10pt]{JHEP3}
\usepackage{graphicx}
\usepackage{epsfig}
\newcommand {\eps}{\epsilon}
\newcommand {\si}{\sigma}
\newcommand {\ga}{\gamma}
\newcommand {\de}{\delta}
\newcommand {\Ga}{\Gamma}

\newcommand {\la}{\lambda}
\newcommand {\al}{\alpha}
\newcommand {\be}{\beta}
\newcommand {\pa}{\partial}
\newcommand {\na}{\nabla}
\newcommand {\fr}{\frac}

\newcommand {\cc}{{\cal C}}
\newcommand {\cd}{{\cal D}}
\newcommand {\cB}{{\cal B}}
\newcommand {\ch}{{\cal H}}
\newcommand {\cu}{{\cal U}}
\newcommand {\cv}{{\cal V}}

\newcommand {\ct}{{\cal T}}
\newcommand {\ce}{{\cal E}}
\newcommand {\beg}{\begin{equation}}
\newcommand {\en}{\end{equation}}
\newcommand {\bega}{\begin{eqnarray}}
\newcommand {\ena}{\end{eqnarray}}
\title{Coupled inflaton and
electromagnetic fields from Gravitoelectromagnetic Inflation with
Lorentz and Feynman gauges.}
\author{Federico Agust\'{\i}n Membiela \\
Departamento de F\'{\i}sica, Facultad de Ciencias Exactas y
Naturales, Universidad Nacional de Mar del Plata, Funes 3350,
(7600) Mar del Plata,
Argentina.\\
Instituto de F\'{\i}sica de Mar del Plata (IFIMAR), Consejo
Nacional de Investigaciones Cient\'{\i}ficas y T\'ecnicas
(CONICET), Argentina. E-mail: \email{membiela@mdp.edu.ar}}
\author{Mauricio Bellini \\
Departamento de F\'{\i}sica, Facultad de Ciencias Exactas y
Naturales, Universidad Nacional de Mar del Plata, Funes 3350,
(7600) Mar del Plata,
Argentina.\\
Instituto de Investigaciones F\'{\i}sicas de Mar del Plata
(IFIMAR), Consejo Nacional de Investigaciones Cient\'{\i}ficas y
T\'ecnicas (CONICET), Argentina.  E-mail:
\email{mbellini@mdp.edu.ar} }

\abstract{Using a semiclassical approach to Gravitoelectromagnetic
Inflation (GEMI), we study the origin and evolution of seminal
inflaton and electromagnetic fields in the early inflationary
universe from a 5D vacuum state. We use simultaneously the Lorentz
and Feynman gauges. Our formalism is naturally not conformal
invariant on the effective 4D de Sitter metric, which make
possible the super adiabatic amplification of electric and
magnetic field modes during the early inflationary epoch of the
universe on cosmological scales. This is the first time that
solutions for the electric field fluctuations are investigated in
a systematic way as embeddings for inflationary models in 4D. An
important and new result here obtained is that the spectrum of the
electric field fluctuations depend with the scale, such that the
spectral index increases quadratically as the scale decreases.}

\begin{document}
\section{Introduction}

The origin of cosmological scales magnetic fields is one of the
most important, fascinating and challenging problems in modern
cosmology. Many scenarios have been proposed to explain them.
Magnetic fields are known to be present on various scales of the
universe\cite{3}. Primordial large-scale magnetic fields may be
present and serve as seeds for the magnetic fields in galaxies and
clusters.

Until recently the most accepted idea for the formation of
large-scale magnetic fields was the exponentiation of a seed field
as suggested by Zeldovich and collaborators long time ago. This
seed mechanism is known as galactic dynamo. However, recent
observations have cast serious doubts on this possibility. There
are many reasons to believe that this mechanism cannot be
universal. This is why the mechanism responsible for the origin of
large-scale magnetic fields is looked in the early universe, more
precisely during inflation\cite{turner}, which should be amplified
through the dynamo mechanism after galaxy formation. In principle,
one should be able to follow the evolution of magnetic fields from
their creation as seed fields through to dynamo phase
characteristic of galaxies. It is believed that magnetic fields
can play an important role in the formation and evolution of
galaxies and their clusters, but are probably not essential to our
understanding of large-scale structure in the universe. However,
an understanding of structure formation is paramount to the
problem of galactic and extragalactic magnetic fields\cite{1,2}.

It is natural to look for the possibility of generating
large-scales magnetic fields during inflation with strength
according with observational data on cosmological scales: $<
10^{-9}$ Gauss\cite{giova}. However, the FRW universe is conformal
flat and the Maxwell theory is conformal invariant, so that
magnetic fields generated at inflation would come vanishingly
small at the end of the inflationary epoch. The possibility to
solve this problem relies in produce non-trivial magnetic fields
in which conformal invariance to be broken.

On the other hand, the five dimensional model is the simplest
extension of General Relativity (GR), and is widely regarded as
the low-energy limit of models with higher dimensions (such as 10D
supersymmetry and 11D supergravity). Modern versions of 5D GR
abandon the cylinder and compactification conditions used in
original Kaluza-Klein (KK) theories, which caused problems with
the cosmological constant and the masses of particles, and
consider a large extra dimension. In particular, the Induced
Matter Theory (IMT) is based on the assumption that ordinary
matter and physical fields that we can observe in our 4D universe
can be geometrically induced from a 5D Ricci-flat metric with a
space-like noncompact extra dimension on which we define a
physical 5D apparent vacuum. The vacuum we shall consider is very
restrictive in the sense that we shall not consider any kind of
charges, matter or currents on the 5D spacetime. In a relativistic
framework, it can be expressed by the 5D null geodesic equations,
which are only valid for massless test particles in 5D. However,
observers that move with frames $U^4 \equiv {d x^4 \over dS}=0$
(described by a constant foliation on the extra dimension), can
see the physics described by the effective 4D energy-momentum
tensor embedded in the 5D apparent vacuum, which is geometrically
described by a 5D Ricci-flat spacetime. From the mathematical
point of view, the Campbell-Magaard theorem\cite{campbell} serves
as a ladder to go between manifolds whose dimensionality differs
by one. This theorem, which is valid in any number of dimensions,
implies that every solution of the 4D Einstein equations with
arbitrary energy-momentum tensor can be embedded, at least
locally, in a solution of the 5D Einstein field equations in
vacuum. Because of this, matter, charge and currents may be 4D
manifestations of the topology of space.

Gravitoelectromagnetic Inflation (GEMI)\cite{gemi} was proposed
recently with the aim to describe, in an unified manner,
electromagnetic, gravitational and the inflaton fields in the
early inflationary universe, from a 5D vacuum. It is known that
conformal invariance must be broken to generate non-trivial
magnetic fields. A very important fact is that in this formalism
conformal invariance is naturally broken. Other conformal symmetry
breaking mechanisms have been proposed so far\cite{tur}. However,
most of these are developed in the Coulomb gauge. In order to
simplify the equations of motion for $A^{\nu}$, in this paper we
use simultaneously the Lorentz and Feynman gauges, to calculate
the electric and magnetic spectral indices for the spectrums of
these fluctuations taking into account the induced currents. The
main contribution of this paper is the study for the spectrum of
the electric field fluctuations in a systematic way as embeddings
for inflationary models in 4D. This topic has been ignored in the
literature.

The paper is organized as follows: in Sect. II we introduce the 5D
vacuum of the fields on a generic 5D Ricci flat metric, to obtain
the equations for the vector fields using simultaneously the
generalized Lorentz and Feynman gauges. Also, we impose a
semiclassical approach to the vector fields. In Sect. III we study
the particular case of a 5D Ricci flat space-time for an extended
de Sitter expansion. In Sect. IV describe the dynamics of the
vector fields on an effective 4D de Sitter space-time, when we
make a static foliation on the noncompact extra dimension, which
is considered as space-like: $\psi=\psi_0$. We develop the
equations of motion for the fields using a particular Lorentz
gauge on the effective 4D de Sitter space-time. After it, we
describe the dynamics of the classical and quantum fields, to
finally calculate the evolution and spectrums of the inflaton,
electric and magnetic fields. The conclusions are developed in the
Sect. V. Finally, we have included two appendixes where we have
developed respectively the details of the calculations for the
modes of the electric field fluctuations, and the spectrum for
these fluctuations.

\section{Vector fields in 5D vacuum}

We begin considering a 5D manifold $\cal{M}$ described by a
symmetric $g_{ab}=  g_{ba}$\footnote{In our conventions latin
indices "a,b,c,..,h" run from $0$ to $4$, greek indices run from
$0$ to $3$ and latin indices "i,j,k,..." run from $1$ to $3$.} 5D
tensor metric. This manifold $\cal M$ is mapped by coordinates
$\{x^a\}$
 \beg
    dS^2=g_{ab}dx^a dx^b.\label{meta}
 \en
From the geometrical point of view, to describe a
 relativistic 5D vacuum, we shall consider that $g_{ab}$ is such
 that the Ricci tensor $R_{ab}=0$, and hence: $G_{ab}=0$.
To describe the system we introduce the action on the manifold
$\cal M$
 \beg
    \mathcal{S}=\int d^5x\sqrt{-g}\left[\fr{^{(5)}\,R}{16\pi
    G}-\fr{1}{4}Q_{bc}Q^{bc}\right],
 \en
where $^{(5)}\,R$ is the 5D scalar curvature on the
five-dimensional metric (\ref{meta}) and $Q^{ab}=F^{ab}-\ga
g^{ab}\na_f A^f$, where the 5D Faraday tensor is $F^{bc}=\na^b
A^c-\na^c A^b=\pa^bA^c-\pa^cA^b$. We shall consider that the
fields $A^b$ are minimally coupled to gravity and free of
interactions, so that the second term in the action is purely
kinetic.

\subsection{Einstein Equations in 5D}

If we minimize the action respect to the metric we will obtain
Einstein Equations in 5D. In this paper we shall
 use a semiclassical approach where the Einstein equations are
 expressed by the homogeneous component of the fields. This slightly
 differs from the one used by \cite{Birrel} in the fact that we
 don't need to renormalize the stress tensor, but at the cost of
 assuming a semiclassical behavior of the fields that rules out the
 dependence with the wavenumber in the calculation of the semiclassical
 Einstein equations
  \beg
    G_{ab}=-8\pi G \,T^{(0)}_{ab},
  \en
where $T^{(0)}_{ab}\equiv \langle T_{ab}(\bar{A}^c)\rangle$.
Notice that we use a semiclassical expansion of the vector fields
\begin{equation}
A^c=\bar{A}^c+\de A^c,
\end{equation}
where the overbar symbolizes the 3D spatially homogeneous
background field consistent with the fixed homogeneous metric and
$\de A^c$ describes the fluctuations with respect to $\bar{A}^c$.
In this sense when we perform the expectation value of the stress
tensor, adopting the ansatz $\langle\de A_c\rangle=0$, only will
appear zero order $T^{(0)}_{ab}$ and the second order
$T^{(2)}_{ab}$ in perturbations terms. The last corresponds to a
feedback term and is related to back-reaction effects, which do
not will be consider in this paper.
 The stress tensor is defined by the fields lagrangian being
 symmetric by definition
 \beg
    T_{bc}=\fr{2}{\sqrt{g}}\left\{\fr{\pa}{\pa g^{bc}}\left(\sqrt{g}{\cal L}_f\right)
                                   -\fr{\pa}{\pa x^e}\left[\fr{\pa}{\pa {g^{bc}}_{,\,e}}
                                   \left(\sqrt{g}{\cal
                                   L}_f\right)\right]\right\}.
 \en
The appearance of variations with respect to derivatives of the
metric is because we are dealing with vector fields whose
covariant derivative operators involve Christoffel symbols (i.e.
ordinary derivatives of the metric).
 In our case the stress tensor reduces to
 \bega
    T_{bc}&=&{F^e}_b F_{ce}+\fr{1}{4}g_{bc}F_{de}F^{de}
    -\la\left\{2{A^e}_{;\,e}\left[A_{(b;\,c)}-\left(2A^{}_{(b}g^{}_{c)h,\,f}+
    g^{}_{hf,\,(b}A^{}_{a)}\right)g^{hf}\right]+\right.\\
    &+& \left. g_{bc}\left[\left(A^e_{,\,ef}+\Ga^e_{de,\,f}A^d+
    \Ga^e_{de}A^d_{,\,f}+2\Ga^e_{ef}A^d_{,\,d}+\fr{3}{2}\Ga^e_{ed}\Ga^a_{af}A^d
    \right)A^f+\fr{1}{2}\left(A^e_{,\,e}\right)^2\right]\right. \nonumber \\
    &+&  \left. g_{bc,\,f}A^f  A^e_{;\,c}\right\},
 \ena
where $\ga^2=\fr{2\la}{5}$.

\subsection{5D dynamics of the fields}

The Euler-Lagrange equations give us the dynamics for $A_b$

 \beg
    \na_f\na^f A^b-R_f^b A^f-(1-\la)\na^b\na_fA^f=0.
 \en
In particular, the choice $\la=1$ is known as Feynman gauge,
somehow equivalent to a Lorentz gauge $\na_fA^f=0$. In this paper
we shall choose simultaneously both conditions. The first one
assures that the balance of each component of any external current
to be null, and the second one is more restrictive, because
assures that each direction (insider or outsider) of each
component of the current to be zero.

It is easy to show that the 5-divergence of the field equation of
motions satisfy the same equation as in a Minkowski space, but
changing ordinary partial derivatives by the covariant derivative
 \beg
    \na^a\na_a\left(\na_f A^f\right)=0.
 \en
Hence, the Lorentz gauge is satisfied for appropriate initial
conditions of $\na_a A^a=0$. With such a choice the field
lagrangian density ${\cal L}_f=-\fr{1}{4}Q^2$ is
 \beg
    {\cal L}'_f=-\fr{1}{2}\na_a A_b \na^a A^b=-\fr{1}{2}\na_\mu A_\nu \na^\mu
    A^\nu-\fr{1}{2}\na_4 A_\nu \na^4 A^\nu-\fr{1}{2}\na_\mu A_4 \na^\mu
    A^4-\fr{1}{2}\na_4 A_4 \na^4 A^4,
 \en
 where $Q^2 = Q^{ab} Q_{ab}$.
For 4D observers living in a hypersurface where the fifth
component of the vector field is normal to it, this extra
dimensional field will manifest separately, like an effective 4D
vector field $A^\nu$ and a 4D scalar field $A^4$. In this sense we
can identify kinetic terms for both, scalar and vector fields, and
the derivatives with respect to the extra dimension may be
interpreted as potential (or dynamical sources) terms joined with
massive terms for each of them.\\

 The stress tensor in this gauge is
 \bega
    T_{ab}=&&-\na_aA_e\na_bA^e-\na_e A_a \na^e A_b- 2
    g_{c(b}A_{a)}\Ga^c_{ef}\na^eA^f +\fr{1}{2}g_{ab}\na_e A_f \na^e
    A^f-\\
    \nonumber&&2\fr{g_{,f}}{g}\left[\na_{(a} A_{b)} A^f+\na^f A_{(b} A_{a)}- A_{(a}\na_{b)}
    A^f\right]-
    \left[\na_{(a} A_{b)} A^f+\na^f A_{(b} A_{a)}- A_{(a}\na_{b)} A^f\right]_{,f}.
 \ena

\section{Special case: 5D generalization of a de Sitter spacetime}

Because we are interested to study a cosmological scenario of
inflation from the context of the theory of Space-Time-Matter, we
shall consider the 5D Riemann-flat metric\cite{LB}
 \beg
    dS^2=\psi^2dN^2-\psi^2
    e^{2N}dr^2-d\psi^2, \label{met1}
 \en
where $N$ is a time-like dimension related to the number of
e-folds, $dr^2=dx^i\de_{ij}dx^j$ is the Euclidean line element in
cartesian coordinates and $\psi$ is the space-like extra
dimension. This metric satisfies the vacuum condition $G^{ab}=0$.

For this 5D metric the field equations, after taking Lorentz
gauge: $\na_a A^a=\pa_N A^0+3A^0+\pa_\psi A^4+4\psi^{-1}A^4+\pa_i
A^i=0$, are

 \begin{eqnarray}
   && \left\{\fr{\pa^2}{\pa N^2}+5\fr{\pa}{\pa N}-e^{-2N}\pa^2_r-
    \psi^2\left[\fr{\pa^2}{\pa\psi^2}+\fr{6}{\psi}\fr{\pa}{\pa \psi}\right]\right\}{A^0}
    +\left[\frac{2}{\psi}\fr{\pa}{\pa N}+2 \frac{\pa}{\pa \psi} + \frac{8}{\psi} \right]\,{A^4}=0,\label{b1}\\
   && \left\{\fr{\pa^2}{\pa N^2}+5\fr{\pa}{\pa N}-e^{-2N}\pa^2_r-
    \psi^2\left[\fr{\pa^2}{\pa\psi^2}+\fr{6}{\psi}\fr{\pa}{\pa \psi}
    \right]\right\}{A^j}-2\pa^j\left(A^0+\fr{A^4}{\psi}\right)=0,\label{b2}\\
   && \left\{\fr{\pa^2}{\pa N^2}+3\fr{\pa}{\pa N}-e^{-2N}\pa^2_r -
    \psi^2\left[\fr{\pa^2}{\pa\psi^2}+\fr{6}{\psi}\fr{\pa}{\pa \psi}+
    \fr{12}{\psi^2}\right]\right\} {A^4}=0.\label{b3}
 \end{eqnarray}
Notice that the (\ref{b3}) is decoupled after applying the Lorentz
gauge. However we see that it is not sufficient to decouple all
the field equations. This is because the non zero connections of
the metric (\ref{met1}) act in a non trivial manner in the vector
fields derivatives. There are 14 non zero Christoffel symbols
 \beg
    \Ga^\mu_{\mu4}=\psi^{-1},\ \ \Ga^i_{i0}=1,\ \
    \Ga^0_{ii}=e^{2N},\ \ \Ga^4_{00}=\psi,\ \
    \Ga^4_{ii}=-\psi e^{2N}. \label{con}
 \en
Therefore, in this Riemann-flat spacetime we obtain the D
'Alambertian of the $A^b$ field
 \beg
    \na_f\na^f A^b=0,
 \en
but, expressed in terms of the ordinary derivatives and the
Christoffel symbols we notice the coupling terms
 \beg
    g^{fh}\left\{\pa_f\pa_h A^b+2\Ga^b_{ef}\pa_h
    A^e+\Ga^b_{he,\,f}A^e-\Ga^e_{fh}\pa_e
    A^b-\Ga^e_{fh}\Ga^b_{ed}A^d+\Ga^b_{ef}\Ga^e_{hd}A^d\right\}=0.
 \en
Notice that in a 5D Minkowskian metric: $dS^2 =  dt^2-dr^2 -
d\psi^2$, the connections vanish and the field equations remain
decoupled after the gauge choice.

\subsection{Dynamics of the 3D spatially isotropic background fields}

We shall combine the field equations of motion for the classical
homogeneous fields with the Einstein Equations, the first ones
reduce to
 \begin{eqnarray}
    \left\{\fr{\pa^2}{\pa N^2}+5\fr{\pa}{\pa N}-
    \psi^2\left[\fr{\pa^2}{\pa\psi^2}+\fr{6}{\psi}\fr{\pa}{\pa \psi}
    \right]\right\}&{\bar{A}^0}&+\left[\frac{2}{\psi}\fr{\pa}{\pa N}
    +2 \frac{\pa}{\pa \psi} + \frac{8}{\psi} \right]\,{\bar{A}^4}=0,\label{c1} \\
    \left\{\fr{\pa^2}{\pa N^2}+5\fr{\pa}{\pa N}-
    \psi^2\left[\fr{\pa^2}{\pa\psi^2}+\fr{6}{\psi}\fr{\pa}{\pa \psi}
    \right]\right\}&{\bar{A}^j}&=0, \label{c2}\\
    \left\{\fr{\pa^2}{\pa N^2}+3\fr{\pa}{\pa N}-
    \psi^2\left[\fr{\pa^2}{\pa\psi^2}+\fr{6}{\psi}\fr{\pa}{\pa \psi}+
    \fr{12}{\psi^2}\right]\right\}&{\bar{A}^4}&=0. \label{c3}
 \end{eqnarray}
Notice that the equation for $\bar A^0$ is the unique coupled.
Furthermore, once obtained ${\bar{A}^4}$, we can describe the
dynamics of ${\bar{A}^0}$ in (\ref{c1}), where ${\bar{A}^4}$
appears as a source.

\section{Effective 4D dynamics of the fields}

Now we consider a static foliation on the 5D metric (\ref{met1}).
The resulting 4D hypersurface after making $\psi=\psi_0$ describes
a de Sitter spacetime. From the relativistic point of view an
observer moving with the penta velocity $U_\psi=0$, will be moving
on a spacetime that describes a de Sitter expansion which has a
scalar curvature $^{(4)} R=12/\psi^2_0=12\,H^2_0$, such that the
Hubble parameter is defined by the foliation $H_0=\psi_0^{-1}$.
Hence, if we consider the coordinate transformations on
(\ref{met1})
 \beg\label{trans}
    t=\psi_0N,\ \ \ R=\psi_0 r,\ \ \ \psi=\psi,
 \en
we then arrive to the Ponce Leon metric\cite{PdL}:
$dS^2=\left(\fr{\psi}{\psi_0}\right)^2\left[dt^2-e^{2t/\psi_0}dR^2\right]-d\psi^2$.
If we foliate $\psi=\psi_0$, we get the effective 4D metric
 \beg
    dS^2\rightarrow ds^2=dt^2-e^{2H_0t}d\vec{R}^2,\label{24}
 \en
which describes a 3D spatially flat, isotropic and homogeneous de
Sitter expanding universe with a constant Hubble parameter $H_0$.

The dynamics of the fields being given by the equations
(\ref{b1}), (\ref{b2}) and (\ref{b3}), evaluated on the foliation
$\psi=\psi_0=1/H_0$, with the transformations (\ref{trans}). In
the following subsections we shall study separately the dynamics
of the classical 3D spatially isotropic fields:
$\bar{A}^{\mu}(t,\psi_0)$ and $\bar{A}^{4}(t,\psi_0)$, and the
fluctuations of these fields: $\delta A^{\mu}(t,\vec{R},\psi_0)$
and $\delta A^{4}(t,\vec{R},\psi_0)$. Notice that now
$\vec{R}\equiv \vec{R}(X^i)$. To describe the dynamics of the
fields we shall impose the effective 4D Lorentz gauge:
$^{(4)}\nabla_{\mu} A^{\mu}=0$. It implies that the 5D Lorentz
gauge with the transformations (\ref{trans}) and evaluated on the
foliation must now be
\begin{equation}\label{ga}
\left.\na_a A^a\right|_{\psi_0} =^{(4)} \nabla_{\mu}
A^{\mu}(t,\vec{R},\psi_0) +\left.\left(\pa_\psi
A^4+4\psi^{-1}A^4\right)\right|_{\psi_0}=0,
\end{equation}
where $^{(4)} \nabla_{\mu}\,A^{\mu}$ denotes the covariant
derivative on the effective 4D metric (\ref{24}). Hence, in order
to the effective 4D Lorentz gauge to be fulfilled, we shall
require
\begin{equation}\label{gau}
\left.\left(\pa_\psi A^4+4\psi^{-1}A^4\right)\right|_{\psi_0}=0.
\end{equation}

\subsection{4D classical field dynamics}\label{cfd}

In order to solve the equations (\ref{c1}), (\ref{c2}) and
(\ref{c3}) on an effective 4D de Sitter metric, we must evaluate
these equations on the particular foliation
$\psi=\psi_0=H^{-1}_0$, $r=R\,\psi_0$ and $N=H_0\,t$. {\bf We
shall identify the effective scalar $A^4$ with the inflaton
field}: $A^4(t,\vec R,\psi_0)\equiv
  \phi(t,\vec R,\psi_0)$ and we shall denote $\bar\phi(t,\psi_0)
\sim \left.\phi_1(N)\,\phi_2(\psi)\right|_{N=H_0
t,\psi=\psi_0=H^{-1}_0}$, as the 3D spatially isotropic and
homogeneous background field. In the same way we state for the
homogeneous component of the vector field the separation $\bar
A^j(t,\psi_0)\sim \left.S^j_{1}(N)S^j_{2}(\psi)\right|_{N=H_0
t,\psi=\psi_0=H^{-1}_0}$, in the next we shall drop the index $j$
to label the functions $S_1(t)$ and $S_2(\psi_0)$. Hence, we
obtain
 \beg
    \bar\phi(t,\psi_0)=e^{-\fr{3}{2}H_0 t}\left(a_1 \, e^{\al H_0 t}+a_2 \, e^{-\al
    H_0 t}\right),\ \ \ \ \al=\fr{3}{2}\sqrt{1-\fr{4m^2}{9}},
 \en
 where we have considered the condition (\ref{gau}), such that
 \beg
\left. -\psi^2 \left[ \frac{\pa^2}{\pa\psi^2} + \frac{3}{\psi}
 \frac{\pa}{\pa\psi} \right] \bar{A}^4\right|_{\psi_0} = m^2
 \,\bar{\phi}(t,\psi_0),
 \en
where $\bar{\phi}$ plays the role of the background inflaton
field.
 Furthermore, the general solution of eq. (\ref{c2}) on the effective 4D metric
 (\ref{24}), is
  \beg
    \bar
A^j(t,\psi_0)\sim S(t)= e^{-\fr{5}{2}H_0 t}\left(c_1 \, e^{\si H_0
t}+c_2 \, e^{-\si H_0 t}\right),\ \ \ \
\si=\fr{5}{2}\sqrt{1-\fr{4\nu^2}{25}}
    \en
where \beg \left. -\psi^2 \left[ \frac{\pa^2}{\pa\psi^2} +
\frac{6}{\psi}
 \frac{\pa}{\pa\psi} \right] \bar{A}^j\right|_{\psi_0} = \nu^2
 \,\bar{A}^j(t,\psi_0).
\en A similar treatment can be done for $\bar{A}^0$, after making
use of the condition (\ref{gau}), the transformations
(\ref{trans}) and the foliation $\psi=\psi_0=1/H_0$. However, the
difference with the other background components of the field
observed in eq. (\ref{c1}) is that $\bar{A}^4\equiv
\bar\phi(t,\psi_0)$ acts as a source of $\bar{A}^0(t,\psi_0)$.

As a particular choice we shall consider a 4D inflationary
universe, where the background fields are $\bar
A^b=\left(0,0,0,0,\bar\phi\right)$, in agreement with a global (de
Sitter) accelerated expansion which is 3D spatially isotropic,
flat and homogeneous. \footnote{One could consider, for instance,
the case when the background field is $\bar A^b=\left(\Phi,\bar
A^1,0,0,0\right)$, that defines an effective homogeneous component
of the electric field. However, we would obtain an anisotropic
component of the stress tensor $T_{10}$, which is not compatible
with our background, spatially flat, homogeneous and { isotropic}
(de Sitter) metric. In general this implies that for the
background fields to satisfy Einstein equations, the components
$\bar A_0;\bar A_1;\bar A_2;\bar A_3$ are highly restricted. In
particular we have the following cases to choose:(i)$\bar A^i=0$,
$\bar A^0=\bar\Phi(t,\psi_0)$ and $\bar A^4=\bar\phi(t,\psi_0)$,
(ii)$\bar A^0=\bar A^4=0$ and $\bar A^i=\bar A_0^i$ constants. In
what follows we shall analyze a particular choice of the first
case (with $\bar A^0=0$), because the other isn't very interesting
in the physical sense.}.
 In this case, the relevant
components of the classical Energy momentum tensor, are
 \bega
    \rho\equiv\langle T^0_0\rangle&=&\fr{1}{2}\dot{\bar\phi}^2+
    \left[\fr{5}{\psi^2} \bar\phi^2+ \fr{1}{2} \bar{\phi}'^2+\fr{2}{\psi}
    \bar\phi \bar{\phi}'\right]_{\psi=\psi_0}, \label{t1}\\
    p\equiv\langle
    -T^i_j\rangle|_{i=j}&=&\fr{1}{2}\dot{\bar\phi}^2-
    \left[\fr{5}{\psi^2} \bar\phi^2+ \fr{1}{2} \bar{\phi}'^2+\fr{2}{\psi}
    \bar\phi \bar{\phi}'\right]_{\psi=\psi_0}, \label{t2}\\
    \langle T^\al_\be\rangle|_{\al\neq\be}&=&0, \label{t3}
 \ena
where the prime denotes the partial derivative with respect to
$\psi$ and dots denote partial derivatives with respect to the
time, which in our case are zero:
$\left.\dot{\bar\phi}\right|_{\psi_0}=0$. Furthermore, from eq.
(\ref{t1}) we can make the following identification for the
background scalar potential:
\begin{equation}
V[\bar\phi]=\left[\fr{5}{\psi^2} \bar\phi^2+ \fr{1}{2}
\bar{\phi}'^2+\fr{2}{\psi} \bar\phi
\bar{\phi}'\right]_{\psi=\psi_0}. \label{pote}
\end{equation}
In our model, the hypersurface $\psi=\psi_0$ defines a de Sitter
expansion of the universe with a Hubble parameter
$H_0=\psi_0^{-1}$. The equation of state for this case is
$p=-\rho=-3/\left(8 \pi G\psi^2_0\right)$. Then, it is easy to see
that the only compatible background solution for the field
evaluated on the hypersurface is the typical de Sitter solution
for a background scalar field:
$\bar\phi(t=N/H_0,\psi_0)=\bar\phi_0$. This means that
\begin{equation}\label{pot}
V\left[\bar\phi(t,\psi_0)\right]=\frac{3 H^2_0}{8\pi G}.
\end{equation}
A particular solution of (\ref{gau}) is
\begin{equation}
\bar{\phi} = \bar{\phi}_0 \left(\frac{\psi_0}{\psi}\right)^4,
\qquad \rightarrow \bar\phi' \equiv \left.\frac{\partial
\bar\phi}{\partial\psi}\right|_{\psi=\psi_0=1/H_0}=- 4 H_0
\bar\phi_0. \label{e2}
\end{equation}
From eqs. (\ref{pote}), (\ref{pot}) and the second in (\ref{e2}),
we obtain
\begin{equation}\label{cam}
\left.\left(\bar\phi\right)^2\right|_{\psi=\psi_0=1/H_0} =
\bar\phi^2_0=\frac{3}{40\,\pi\,G},
\end{equation}
such that replacing (\ref{cam}) in the second equation of
(\ref{e2}), we obtain
\begin{equation}
(\bar\phi')^2\equiv \left(\left.\frac{\partial
\bar\phi}{\partial\psi}\right|_{\psi=\psi_0=1/H_0}\right)^2=
\frac{6 H^2_0}{5\pi G}.
\end{equation}
It is easy to see by inspection in (\ref{c1}) that
$\bar\phi(t,\psi)$ is a constant of $N$. In other words, the
unique origin of the effective 4D potential energy density
(\ref{pot}) related to the background inflaton field is the
$\psi$-dependence of $\bar\phi(N,\psi)$.

\subsection{4D Field fluctuations}

Here we consider equations (\ref{b1}), (\ref{b2}) and (\ref{b3}) to
search for possible electromagnetic fields generated through this
model. In Sect. (\ref{cfd}) we've seen that the Einstein equations
for the background fields exclude any possibility of spatially
homogeneous electromagnetic fields.

The equation for the effective scalar $\delta
A^4(t,\vec{R},\psi_0)$ on the effective hypersurface (\ref{24}) is
decoupled from the dynamics of the 4-vector. In contrast, the
equations for $\de A^0(t,\vec{R},\psi_0)$ and $\de
A^i(t,\vec{R},\psi_0)$ remain coupled. By the use of our 5D
Lorentz gauge evaluated on the foliation $\psi=\psi_0=H^{-1}_0$:
$\left. \nabla_{a} \,A^a\right|_{\psi_0=H^{-1}_0}=0$, we can
express the inhomogeneous term for $\de A^0$ as only a function of
$\de A^4$. The solution will involve both, homogeneous and
inhomogeneous parts. Once obtained $\delta A^0$ and $\delta A^4$,
we can finally search solutions for the components $\de A^j$.
These total solutions are necessary to deduce the effective
electric fields. In contrast, as we previously said, the equation
of motion for pure magnetic fields may be obtained by just
applying the curl in the 3-space to equation (\ref{b2}). The last
term in (\ref{b2}) vanishes because is a 3-gradient, and so
magnetic fields equations are decoupled. To quantize the field
fluctuations on the effective 4D de Sitter spacetime (\ref{24}),
we shall consider the equations (\ref{b1}), (\ref{b2}) and
(\ref{b3}), with condition (\ref{gau}), the transformations
(\ref{trans}) and the foliation $\psi=\psi_0=1/H_0$. The equal
time canonical relations are

 \beg\label{com1}
  \left.\left[ \de A_i(t,\vec{R},\psi_0),
  \Pi^j(t,\vec{R'},\psi_0)\right]\right|_{\psi_0=1/H_0} = -i\,
  g^j_i e^{-3H_0t}\,\delta^{(3)}(\vec{R} - \vec{R'}),
 \en
where
$g^{ij}$ are the space-like components of the tensor metric in
(\ref{24}) and $\delta^{(3)}(\vec{R} - \vec{R'})$ is the 3D
Dirac's function. Furthermore, the canonical momentum is given by
the electric field $\Pi^j \equiv E^j= \nabla^j A^0 - \nabla^0
A^j$. The equations (\ref{b1}), (\ref{b2}) and (\ref{b3}) with the
transformations (\ref{trans}) can be evaluated on the foliation
$\psi=\psi_0=1/H_0$ to give the dynamics on the effective 4D
spacetime (\ref{24}). If we take into account the conditions
(\ref{gau}), the effective 4D dynamics of the fluctuations
describe an effective 4D Lorentz gauge, so that
\begin{eqnarray}
&&\frac{\partial^2\delta A^0}{\partial t^2} + 5 H_0
\frac{\partial\delta A^0}{\partial t} - H^2_0 e^{-2H_0 t}
\partial^2_R \delta A^0 + \nu^2H^2_0 \delta A^0= -2 H^2_0
\frac{\partial\delta
\phi}{\partial t}, \label{ff1} \\
&&\frac{\partial^2\delta A^j}{\partial t^2} + 5 H_0
\frac{\partial\delta A^j}{\partial t} - H^2_0 e^{-2H_0 t}
\partial^2_R \delta A^j + \nu^2H^2_0 \delta A^j= 2 H^2_0
\pa^j\left(\delta A^0+ H_0 \delta\phi\right),
\label{f2} \\
&&\frac{\partial^2\delta\phi}{\partial t^2} + 3 H_0
\frac{\partial\delta\phi}{\partial t} - H^2_0 e^{-2H_0 t}
\partial^2_R \delta\phi + m^2 H^2_0 \delta\phi= 0. \label{f3}
\end{eqnarray}
describe the 4D dynamics of the fluctuations. A very important
fact is that the electromagnetic field fluctuations $\delta
A^{\mu}$ obey a Proca equation with sources.

 The expansion of the free
field in temporal modes is
 \beg
    \de A^\mu(t,\vec{R},\psi_0)=\int \fr{d^3 K}{(2\pi)^3}\sum_{\la=1}^3
    \varepsilon^\mu(\vec{K},\la)\left(a_{(\vec{K},\la)}e^{-i\vec{K}\cdot\vec{R}}W(K,t,\psi_0)+
    a^\dagger_{(\vec{K},\la)}e^{i\vec{K}\cdot\vec{R}}W^\star(K,t,\psi_0)\right),
    \en
The equation of motion for the temporal modes $W(\vec{K},t,\psi_0)$
of the free contravariant fluctuations $\de A^\mu$ is
 \begin{equation}
\left\{\frac{\partial^2 }{\partial t^2} + 5 H_0
\frac{\partial}{\partial t} + \left[ K^2 e^{-2H_0 t} +
\nu^2H^2_0\right]\right\} W= 0, \label{g1}
 \end{equation}
where $\vec K= H_0 \, \vec k$ ($k$ is a dimensionless wavenumber).
Furthermore, $\varepsilon^{\mu}(\vec{k},\lambda)$ are the
polarizations \footnote{parenthesis denotes that sum do no run over
these indices.}, such that in the Lorentz gauge the following
expression holds:
\begin{equation}
\sum_{\lambda=1}^{3}
\varepsilon_{\alpha}(\vec{k},\lambda)\,\varepsilon_{\beta}(\vec{k},\lambda)
= -\left(g_{\al\be}-\fr{H_0^2}{m^2_{eff}}k_\al k_\be\right),
\end{equation}
where we have introduced the effective mass
$m^2_{eff}=H_0^2(\nu^2-\fr{25}{4})$ of the redefined temporal
modes ${\cu}_K(t)=e^{5H_0t/2}W(K,t,\psi_0)$, that obey the
harmonic equation $\ddot\cu_K+\omega^2_K(t)\cu_K=0$. The time
dependent frequency is defined by the relation $K_\mu
K^\mu=m^2_{eff}$.
 \beg
    \omega^2_K(t)=\left[m^2_{eff}+(e^{-H_0t} K)^2\right].
 \en
 Modes with $\omega^2_K >0$ are stable, but those with $\omega^2_K
 <0$ [i.e., with $k < \left(25/4 - \nu^2 \right)^{1/2} e^{H_0 t}$], are unstable.
In the small wavelength limit these behave like plane waves in
Minkowski space. Furthermore, the annihilation and creation
operators $a_{(K,\lambda)}$ and $a^{\dagger}_{(K,\lambda)}$,
comply with the commutation relations
\begin{equation}
\left[a_{(\vec{K},\lambda)},
a^{\dagger}_{(\vec{K}',\lambda')}\right] = \left(2\pi\right)^3
g_{\lambda\lambda'}\,\delta^{(3)}(\vec{K} - \vec{K}').
\end{equation}
The solutions for the temporal modes is
 \begin{equation}
W(K,t,\psi_0)
=e^{-5H_0t/2}\left\{c_1\ch^{(1)}_\sigma\left[x(t)\right]+c_2\ch^{(2)}_\sigma\left[x(t)\right]\right\},\
\ \ \ \
                               \sigma=\sqrt{\fr{25}{4}-\nu^2},\ \ \ \ \ x(t)={K\over H_0}\, e^{-H_0t}.\\
 \end{equation}
 where $\ch^{(1,2)}_\sigma[x(t)]$ are the first and second kind Hankel
 functions respectively. We can also obtain the temporal modes for the covariant $\de
A_\mu$ which are related to the contravariant ones:
$\ct_K(t)=e^{2H_0t}\,W(K,t,\psi_0)$. The commutation relations
(\ref{com1}) yield the following conditions over these modes
 \bega
    \fr{k^ik_j}{m_{eff}^2}2ik^0\ct_K\ct_K^\star+\left(\fr{k^ik_j}{m_{eff}^2}-\de_j^i\right)
    (\ct_K\dot\ct_K^\star-\ct_K^\star\dot\ct_K)=i\de^i_je^{-H_0t}
 \ena
From these relation we can deduce the following apparently
independent equations
 \bega
    \ct_K\dot\ct_K^\star-\ct_K^\star\dot\ct_K&=&-i e^{-H_0t}, \label{e47}\\
    \ct_K\ct_K^\star&=&\fr{e^{-H_0t}}{2w_K(t)}, \label{e48}
 \ena
which are only valid on (short) wavelength modes for which
$\omega^2_K >0$. Equations (\ref{e47}) and (\ref{e48}) give us the
normalization conditions for the modes of $\delta A_{\mu}$. On the
other hand, these modes are unstable on cosmological scales:
$\omega^2_K <0$, and the expression (\ref{e47}) tends to zero. To
apply these conditions we take the very small wavelength limit for
the Hankel Functions $x(t)\gg|\sigma^2-\fr{1}{4}|$. These means
that $K/H_0\,e^{^-H_0t}\gg m_{eff}^2$, so that $w_K(t)\simeq K
e^{-H_0t}$. In this limit the conditions (\ref{e47}) and
(\ref{e48}) become dependent one of the another, since
\begin{displaymath}
\left.\ct_K\ct_K^\star\right|_{UV}
=\left.\fr{e^{-H_0t}}{2w_K(t)}\right|_{UV} \simeq \frac{1}{2 \,K}.
\end{displaymath}
Letting us choose $c_1=0$ (Bunch-Davies vacuum), the solution for
the modes is
 \beg
    \ct_K(t)=e^{-\fr{1}{2}H_0t}\sqrt{\fr{\pi}{4H_0}}\,\ch^{(2)}_\sigma[x(t)],
 \en

\subsubsection{4D electromagnetic fluctuations}

The electric field for a observer in 4D is defined by its
4-velocity $E_\nu=F_{\nu\la}u^\la$. If we choose the particular
co-moving frame $u^\nu=\left[(H_0\psi_0)^{-1},\vec{0}\right]$, we
obtain
 \bega
    \nonumber E_0&=&0,\\
              E_i&=& \fr{\pa}{\pa
              X^i} \de A^0- e^{2H_0 t} \fr{\pa }{\pa
              t} \de A^i-2 H_0 \,e^{2H_0 t} \de A^i.
 \ena
The magnetic fields are defined by
$B_\nu=\fr{1}{2}\eps_{\nu\la\al\be}u^\la F^{\al\be}$, where
$\eps_{\nu\la\al\be}=\sqrt{\left|{}^{(4)}g\right|}{\cal
A}_{\nu\la\al\be}$ is the totally antisymmetric Levi-Civita tensor
and ${\cal A}_{\nu\la\al\be}$ is a totally antisymmetric symbol
with ${\cal A}_{0123}=-1$. Then for a co-moving observer we will
have a magnetic field,
 \bega
\nonumber    B_0&=&0,\\
\nonumber    B_j&=&\fr{\sqrt{\left|{}^{(4)}g\right|}}{2}\,{\cal
                    A}_{j0kl}\,u^0\,F^{kl}.
 \ena
From the last expression we can arrive to another that will be
useful to obtain an equation of motion for the magnetic fields, we
first define the Levi-Civita symbol in the 3-flat space using the
co-moving frame: $\eps_{jkl}={\cal A}_{j0kl}$ (we note that
$\eps_{123}=1$). Hence
 \beg
    B_j=\sqrt{\left|{}^{(4)}g\right|}
    g^{kk'}u^0\eps_{jkl}\pa_{k'}A^l.
 \en
For our particular case we obtain
 \beg
  e^{-H_0 t}\, B_j=\left[\delta^{k k'} \eps_{jkl}\, \pa_{k'}\right]
   A^l.
 \en
The differential operator between square brackets commutes with
the one applied to $A^j$ in the equation (\ref{b2}), so that in
the equation of motion for ${\cal B}_j=e^{-H_0t}B_j$ there will be
no sources. We can express the field in Fourier components of the
$\de A^j$ field
 \bega
    \cB^j\left(t,\vec{R},\psi_0\right)=\int
    \fr{d^3K}{(2\pi)^3}\sum_{\la=1}^{3}\varepsilon_{l}(\vec{K},\la)\eps^{jnl}\left[ a_{(\vec{K},\la)}
    {\cv}_{n}(K,t,\psi_0)\,
    e^{i\vec{K}\cdot\vec{R}} + a^{\dagger}_{(\vec{K},\la)} \cv^\star_{n}(K,t,\psi_0)\,e^{-i\vec{K}\cdot\vec{R}}\right].
 \ena
Here $\cv_{j}(K,t,\psi_0)=-iK_j\, W(K,t,\psi_0)$ are the temporal
modes with their complex conjugate
$\cv^\star_{j}(K,t,\psi_0)=iK_j\,W^\star(K,t,\psi_0)$. We perform
the vacuum expectation value of the B-fields quadratic amplitude,
defined by the invariant product $\langle B^2\rangle\equiv
\langle0|B^\al B_\al|0\rangle$. For comoving observers $B^0=0$ and
so we have $B^2=B^j\,B_j=e^{-2H_0t}\sum_j{B_j}^2=\sum_j{\cB_j}^2$.
Then
 \beg
    \langle B^2\rangle=\int \fr{d^3K}{(2\pi)^3}(2e^{2H_0t}K^2)W(K,t,\psi_0)
    W^\star(K,t,\psi_0).
 \en
We will cut the above integral up to wavelengths that remain well
outside the horizon wavenumber $k_H=\si e^{H_0t}$. In this limit
we use the asymptotic limit of the Hankel functions for the long
wavelength limit $k\,e^{-H_0t}\ll\sqrt{\sigma+1}$. The power
spectra is then
 \beg
    {\cal
    P}_B(k)=\fr{2^{2\sigma}\Ga^2(\sigma)H_0^4}{4\pi^3}\,e^{(2\sigma-3)H_0t}\,k^{5-2\sigma},
 \en
if we ask for an almost scale invariant spectrum, then
$\sigma=\fr{5}{2}+\eta,\ \ \eta=-\fr{\nu^2}{5}$ and $|\nu^2|\ll
1$. The quadratic amplitude is then
 \beg
    \langle
    B^2\rangle=\fr{45H_0^4}{4\pi^2\nu^2}e^{2H_0t}\left(\fr{5\theta}{2}\right)^{-2\eta},
 \en
where $\theta \ll 1$ is a control parameter, such that we stay
with super Hubble wavelenghts: $k < \theta\, k_H$.

Using only the homogeneous solutions of the equations (\ref{ff1})
and (\ref{f2}) we can deduce their contribution for electric
fields on the infrared (IR) sector, we obtain for comoving
observers $\langle E^2\rangle_{IR}=\langle
E_A^2+E_B^2+E_C^2\rangle_{IR}$, where
 \bega
\langle E_A^2\rangle_{IR} &\simeq&-H_0^5\fr{e^{-4H_0t}}{\left(\nu^2-\fr{25}{4}\right)}\,\int_0^{\theta k_H}\fr{dk}{2\pi^2}\,k^6|\ct_k|^2,\\
\langle E_B^2\rangle_{IR} &\simeq&-H_0^5\,e^{-2H_0t}\,\int_0^{\theta k_H}\fr{dk}{2\pi^2}\,\left(3\,e^{2H_0t}+\fr{H_0^2k^2}{m_{eff}^2}\right)|\dot\ct_k|^2,\\
\langle E_C^2\rangle_{IR} &\simeq & H_0^5\,e^{-2H_0t}\,
\int_0^{\theta
k_H}\fr{dk}{2\pi^2}\sum_j\fr{H_0^2k_0k_j}{m_{eff}^2}(-iH_0k_j)\left(\ct_k\dot\ct_k^\star-\ct_k^\star\dot\ct_k\right).
 \ena
The corresponding power spectrums are
 \bega
    {\cal P}_{E_A}(k)&=&\fr{2^{2\sigma}\Ga^2(\sigma)H_0^4}{8\si^2\pi^3}\,e^{(2\sigma-5)H_0t}\,k^{7-2\sigma},\\
    {\cal P}_{E_B}(k)&=&\fr{2^{2\sigma}\Ga^2(\sigma)H_0^4}{8\pi^3}(\si^2+\si+1/4)\left(3 e^{(-1+2\si)H_0t}k^{3-2\si}
    +\si^{-2}e^{2\si H_0t}k^{5-2\si}\right),\\
    {\cal P}_{E_C}(k)&=&0
  \ena
The last goes to zero in cosmological scales since it is
proportional to the wronskian (\ref{e47}). If we choose
$\sigma=\fr{5}{2}+\eta,\ \ \eta=-\fr{\nu^2}{5}$ and $|\nu^2|\ll1$,
we get
 \bega
  \langle  E_A^2 \rangle_{IR} &\simeq&\left(\fr{3}{2\pi}\right)^2\,H_0^4\,e^{2H_0t}\theta^2,\\
    \langle E_B^2 \rangle_{IR} &\simeq& \left(\fr{9}{5\pi}\right)^2\,H_0^4\,e^{2H_0t}
    \left[3\theta^{-2}+\fr{4}{25}\theta^{-2\eta}\right],\\
  \langle E_C^2 \rangle_{IR} &\simeq& 0,
 \ena
on cosmological scales. Notice that $\langle E^2 \rangle$ is not
scale invariant for a scale invariant magnetic field. Then we can
say that on very large scales the amplitude of electromagnetic
fields are
 \bega
    \left<B^2\right>^{1/2}_{IR}\simeq\fr{3\sqrt5}{2\pi\nu}H_0^2
    e^{H_0t}\left(\fr{5\theta}{2}\right)^{\nu^2/5},\ \ \ \ \
    \left<E^2\right>^{1/2}_{IR}\simeq\fr{3^{5/2}}{5\pi}H_0^2e^{H_0t}\theta^{-1},
    \label{magg}
 \ena
 which are related to comoving observers. During inflation, the strength of the
 magnetic field in a physical frame is
 \begin{equation}
 \left< B^2_{phys}\right>^{1/2} \sim e^{-2H_0
 t}\,\left<B^2\right>^{1/2}_{IR},
 \end{equation}
 where $\left<B^2\right>^{1/2}_{IR}$ is given by the first equation
in (\ref{magg}). At the end of inflation
 (i.e., for $t=t_e$), the size of the horizon was close to $3.6
 \times 10^{-6}$ cm. It has suffered an exponential growth $\simeq 4.4\times 10^{26}$ (we suppose that the
 number of e-folds is $N_e =63$), from its initial value at Planckian scales.
 Hence, we can make an estimation
 for the strength magnetic fields at the end of inflation $\left<
\left( B^{(0)}_{phys}\right)^2\right>^{1/2}_{IR}$ on cosmological
scales
\begin{equation}
\left< \left( B^{(0)}_{phys}\right)^2\right>^{1/2}_{IR} \simeq
\frac{3 \sqrt{5}}{9 \pi \nu} H^2_0 \, \left(\frac{5 \theta}{2}
\right)^{\nu^2/5} \times 10^{-26}.
\end{equation}
If we take $H_0 = 10^{-9}\,{\rm M_p}$, it holds $\simeq
10^{-44}\,{\rm M^2_p} \simeq 10^{16}\,{\rm Gauss}$, where ${\rm
M^2_p} \simeq 0.223\times 10^{60}\,\,{\rm Gauss}$ (1 Gauss
$\simeq$ $0.6476 \times 10^{-21}\,\,{\rm GeV^2}$)\footnote{In all
the paper we consider natural units: $\hbar=c=1$.}. However, must
be noted that this value is very sensitive to the number of
e-folds suffered during inflation.

On the other hand, the present day size of the universe is of the
order of $10^{28}$ cm. (for $t=t_0=1.26 \times 10^{11}\, G^{1/2}$
). We shall suppose that, after inflation $\left<
B^2_{phys}\right>^{1/2}$ decreases adiabatically as $a^{-2}$:
$\left(a(t_0)/a(t_e)\right)^{-2} \simeq 10^{-68}$, so that the
present day value for residual magnetic fields should be of the
order of $\simeq 10^{-52}$ Gauss. Of course, in this estimation is
omitted any possible mechanism for the amplification of these
magnetic fields\cite{amp}, which could be taken into account.

\subsubsection{4D inflaton fluctuations}

For the fluctuations of the inflaton field we can make a similar
treatment. The Fourier expansion is
 \beg
    \de \phi\left(t,\vec{R},\psi_0\right)=\int
    \fr{d^3K}{(2\pi)^3}
\left[ \alpha_{(\vec{K})} \phi(K,t,\psi_0)\,
e^{i\vec{K}\cdot\vec{R}} + \alpha^{\dagger}_{(\vec{K})}
\phi^*(K,t,\psi_0)\,e^{-i\vec{K}\cdot\vec{R}}\right],
 \en
such that the annihilation and creation operators
$\alpha_{(K,\lambda)}$ and $\alpha^{\dagger}_{(K,\lambda)}$,
comply with the commutation relations
\begin{equation}
\left[\alpha_{(\vec{K})}, \alpha^{\dagger}_{(\vec{K}')}\right] =
\left(2\pi\right)^3 \,\delta^{(3)}(\vec{K} - \vec{K}').
\end{equation}
The solutions for the modes $\phi(K,t,\psi_0)$, are
 \begin{equation}
    \phi(K,t,\psi_0)  = e^{-3H_0 t/2}\,\left\{c_1\,J_\mu\left[x(t)\right]
    +c_2\,Y_\mu\left[x(t)\right]\right\},\ \ \ \ \ \mu=\sqrt{\fr{9}{4}-m^2}.
 \end{equation}
The nearly invariant spectrum of the scalar perturbations is
obtained for small values of the effective mass: $|m^2|\ll 1$.
After normalization of the modes, we obtain the standard result
(see, for instance\cite{bcms}), on cosmological scales
 \beg
   \phi(k,t,\psi_0)=\sqrt{\fr{\pi}{4H_0^3}}\ch^{(2)}_\mu[k^{-H_0t}],\ \
   \ \ \ \ \ \mu=\sqrt{\fr{9}{4}-m^2}
 \en
with amplitude
\begin{equation}
\left< \delta\phi^2\right>_{IR} \simeq
\frac{\Gamma^2(\mu)}{\pi^3(3-2\mu)}\left(\frac{2
}{\theta\mu}\right)^{2\mu-3}\,  H^2_0,
\end{equation}
which is divergent for an exactly scale invariant power spectrum
corresponding to a null value of the inflaton field mass $m$.

\subsection{Effective 4D electromagnetic fluctuations with sources included}

In this section we shall find inhomogeneous solutions for the
Fourier components of the fields, we have noted previously that
magnetic fields are only generated through homogeneous solutions.
Instead, electric fields are affected by the coupled dynamics of
the equations of the model. This couplings come from the 5D
background, because some connections in the 5D metric are not null
[see (\ref{con})].
 The equations (including sources) (\ref{ff1})
and (\ref{f2}) may be written as
 \beg \label{sour}
    \left\{\fr{\pa^2}{\pa t^2}+5H_0\fr{\pa}{\pa
    t}+e^{-2H_0t}\,K^2+\nu^2 H_0^2\right\}X^{\mu}={\cal
    F}^{\mu},
 \en
with an inhomogeneous solution
 \beg
    X^{\mu}(t,k,\psi_0)=\fr{\pi e^{-\fr{5}{2}H_0t}}{2H_0\sin(\si\pi)}\int^{t} d\tau {\cal
    F}^{\mu}(\tau)e^{\fr{5}{2}H_0\tau}\left\{J_{\si}[x(\tau)]J_{-\si}[x(t)]
    -J_{\si}[x(t)]J_{-\si}[x(\tau)]\right\}.\label{eu1}
 \en
where ${\cal F}^{\mu}(t)$ are different sources for each of the
equations.
 Using the identities for the Bessel functions and their derivatives we arrive to the
following expression for Fourier transform of the source term of
(\ref{ff1}):
 \beg
{\cal F}^{0}(t)=-\sqrt{\pi
H_0^3}e^{-\fr{3}{2}H_0t}\left\{(3/2-\mu)\ch^{(2)}_\mu[x(t)]+x(t)\ch^{(2)}_{\mu-1}[x(t)]\right\}.
 \en
Notice that the sources ${\cal F}^{\mu}$ were omitted in a
previous treatment\cite{mb}. However, such that sources should be
important, mainly for electromagnetic fields.

Once known the solutions $X^j$ [see appendix(\ref{apen})], we can
define the Fourier components of the electric field:
 \beg\label{fou elec}
E_{(X)}^j(K,t)=-i K^j\, X^0(K,t)-\dot X_1^j(K,t)-\dot
X_2^j(K,t)-\dot X_3^j(K,t).
 \en
 Here, the suffix $(X)$ means that we are dealing with the electric field calculated
only with the inhomogeneous contribution of de modes in $A^\nu$:
$X^\nu(k,t,\psi_0)$.

The amplitude of these fields on cosmological scales [i.e., the
infrared (IR) sector], is given by the expression
 \beg\label{elec1}
\left<E_{(X)}^2\right>=\int_0^{\theta\si e^{H_0t}}
\fr{d^3K}{(2\pi)^3} \sum_j E_{(X)}^j(K,t){E_{(X)}^j}^\star(K,t),
 \en
which has a power spectrum
 \beg
    {\cal P}_k(t)=\fr{H_0^3k^3}{2\pi^2}\sum_j|E^j_{(X)}|^2.
 \en
using the solutions we may write the power spectrum in the
aproximate form [see appendix (\ref{apb})]
 \beg
{\cal P}_k(t)\simeq k^2\sum_{q=0}^\infty a_q
    (ke^{-H_0t})^{\be_0+q}
 \en
 the dominant contribution for the electric field comes from the smaller spectral power with
 $q=0$, setting like previously $\mu=3/2+\eps$ and $\si=5/2+\eta$ we
 obtain
 \beg
    \left< E_{(0)}^2 \right> \sim
    e^{2H_0t}\fr{H_0^4}{(10\pi)^2}\fr{\eps^2}{(\eps-\eta)^4}\theta^2.
 \en
The first correction comes from $a_2$ coefficient, since there is
no $a_1$. When we consider this term, we have no longer scale
independence of the spectral index, and
 \beg
    \de n_k\simeq \fr{a_2}{a_0}\fr{{(ke^{-H_0t})^2}}{\ln k}.
 \en
We shall only write an approximated expression for $a_2$. As it is
shown in (\ref{apb9}), $X^j_3$ dominates, so after considering
only this contribution, one obtains
 \beg
     a_2=-\fr{H_0^4 2^2}{3\pi^2 5^4}\fr{\eps^2}{(\eps-\eta)^4},
 \en
where $\fr{a_2}{a_0}\simeq -\fr{2}{3}$. The first correction due
to the inhomogeneous contribution of the modes of $A^{\nu}$ to the
electric field amplitude is
 \beg
  \left< E_{(1)}^2\right> \sim
  \left<E_{(0)}^2\right>\,\left(1-\fr{25}{12}\theta^2\right),
 \en
where we remember that $\theta = k/k_H \ll 1$, $k_H=\sigma e^{H_0
t}$ being the wavenumber related to the Hubble radius in a
comoving frame. In a physical frame we obtain
$\left<E^2\right>^{1/2} \sim a^2 \,\left<E_{phys}^2\right>^{1/2}$
and the energy density: $\rho \sim a^{4} \rho_{phys}$. Notice that
the energy density related to the electric fields at the end of
inflation ($H_0 \simeq 10^{-9}\, M_p$), is very small with respect
to the background inflaton energy density:
\begin{displaymath}
\frac{\left< E_{(1)}^2\right>_{phys}}{\rho} \simeq \frac{\left<
E_{(1)}^2\right>_{phys}}{V} \sim \left[ 10^{-2}
\left(\frac{H_0}{M_p}\right)\right]^2  \sim
 10^{-22},
\end{displaymath}
so that back-reaction effects due to the electric fields are
really negligible during inflation.\\

The figure (\ref{f1}) shows $\delta n_k$ as a function of
$\theta$. Notice that $\delta n_k$ decreases almost quadratically
as the wavelength decreases. When the horizon entry (i.e., after
inflation when $\theta=\theta_*=k_*/k_{H_*}=1$), the value of
$\delta n_k$ is close to $\delta n_{k_*}\simeq -0.035$. However,
during inflation the cosmological scales wavenumbers are $k/k_H <
10^{-3}$ that corresponds with $\delta n_{k} > -10^{-6}$. Notice
that we have taken into account the value $k_{H_*} = \sigma
e^{60}$ as the wavenumber related to the horizon wavelength when,
after inflation, the horizon entry.

\section{Final Comments}

We have shown how primordial electromagnetic fields and inflaton
fluctuations can be generated jointly during inflation using a
semiclassical approach to GEMI. We have used simultaneously the
Lorentz and the Feynman gauges. The first one assures that the
balance of each component of any external current to be null, and
the second one is more restrictive, because assures that each
direction (insider or outsider) of each component of the current
to be zero. This is done with the aim to assure a 5D vacuum on the
5D Ricci flat metric (\ref{met1}). In correspondence with this
concept of vacuum, we have defined a 5D totally kinetic Lagrangian
density ${\cal L}_f=-\fr{1}{4}Q^2$, which is totally absent of any
kind of interactions.

One of the important facts is that our formalism is naturally not
conformal invariant on the effective 4D metric (\ref{24}), which
make possible the super adiabatic amplification of the modes of
the electromagnetic fields during inflation in a comoving frame on
cosmological (super Hubble) scales.

In this paper we have analyzed the simplest nontrivial
configuration field:
$\bar{A}^b=\left[0,0,0,0,\bar\phi(t,\psi_0)\right]$. For this
configuration of the background fields, the background inflaton
field must be a constant on the metric (\ref{24}) to satisfy the
Einstein background equations in a de Sitter expansion:
$\bar\phi(t,\psi_0)=\bar\phi_0$. Then, in the model here
developed, the expansion of the universe is driven by the
background inflaton field $\bar\phi_0$ and background
electromagnetic fields are excluded to preserve global isotropy.
Notice that back reaction effects are not included, because the EM
field does not contribute to the background expansion of the
universe\cite{Durrer}, but however comes into play an important
role at the perturbative level as vectorial metric fluctuations
which are the geometrical reaction to the vector physical
fields\cite{du}.

To describe the effective 4D dynamics of the fields, we impose the
effective 4D Lorentz gauge $^{(4)}\nabla_{\mu} A^{\mu}=0$, given
simultaneously by conditions (\ref{ga}) and (\ref{gau}).
Therefore, the origin of the generation of the seed of
electromagnetic fields and the inflaton field fluctuations during
inflation can be jointly studied. The dynamics of $\delta A^{\mu}$
on the effective 4D metric (\ref{24}) obey a Proca equation with
sources where the effective mass of the electromagnetic field
fluctuations is induced by the foliation $\psi=\psi_0=1/H_0$. From
the point of view of a relativistic observer this foliation imply
that the component of the penta-velocity $U^{\psi} ={d \psi\over d
S}=0$.

We have obtained that for small values of $\nu$ a nearly
scale-invariant long wavelengths power spectrum for $\left<
B^2\right>^{1/2}$, which grows as $a$ during inflation on a comoving
frame. However, on a physical frame it suffer a super adiabatical
evolution, so that at the end of inflation is of the order of
$10^{16}$ Gauss. After inflation we have supposed that the field
evolves adiabatically as $a^{-2}$, to estimate the present day
values on cosmological scales (for a physical frame): $\left.\langle
B^2_{phys}\rangle ^{1/2}\right|_{{\rm Now}} \simeq 10^{-52}$
Gauss\cite{ul}. On the other hand, the dominant terms in the
amplitude of $\left< E^2\right>$ grows as $a^2$ on a comoving frame,
and has a scale dependent power spectrum with a spectral index
$n_k\simeq 3 +\delta n_k$. This is the main result of this paper.
This scale dependence is described by $\delta n_k$, which decreases
quadratically as the scale decreases. In the limit case where
$k\rightarrow 0$ (very large scales), one finds that $\delta n_{k=0}
\rightarrow 0$. Finally, in what respect to the inflaton field
fluctuations $\left< \delta\phi^2\right>$, we obtain that they are
nearly scale invariant on cosmological scales, and the amplitude is
freezed in agreement with the predictions of standard 4D inflation.

\acknowledgments{ The authors acknowledge CONICET and UNMdP
(Argentina) for financial support.}

\begin{appendix}

\section{The modes of the electric field}\label{apen}

In order to solve the integrate in (\ref{eu1}) we express all the
Hankel functions in terms of the first kind Bessel functions
$J_\al[x(t)]$, and $J_{-\al}[x(t)]$
 \bega
    \ch^{(1)}_\al(x)=\fr{J_{-\al(x)}-e^{-\al\pi i}J_\al(x)}{i\sin(\si\pi)},\\
    \ch^{(2)}_\al(x)=\fr{J_{-\al(x)}-e^{\al\pi
    i}J_\al(x)}{-i\sin(\si\pi)},
 \ena
and then we expand them in their series representation
 \beg
    J_\al[x(t)]=\sum_{m=0}^\infty\fr{(-1)^{m}}{m!\Ga(1+m+\al)}\left(\fr{x(t)}{2}\right)^{2m+\al},
 \en
such that the product identity is\cite{abra}
 \beg
    (J_\al J_\be)[x(t)]=\sum_{m=0}^\infty\fr{(-1)^m}{m!}\fr{\Ga(1+2m+\al+\be)}{\Ga(1+m+\al)
                           \Ga(1+m+\be)\Ga(1+m+\al+\be)}\left(\fr{x(t)}{2}\right)^{2m+\al+\be}.
 \en
The terms included in expression (\ref{eu1}) are of the form,
 \bega
    &&J_\ga(x(t))\int dt e^{H_0 t}(J_\al J_\be)[x(t)]=\fr{k}{2H_0}
    \sum_{m,n}^\infty\fr{(-1)^{m+n}}{m!n!}\cc^{mn}_{\al,\be,\ga}\left(\fr{x(t)}{2}\right)^{2(m+n)-1+\al+\be+\ga},\\
    &&J_\ga(x(t))\int dt (J_\al J_\be)[x(t)]=\fr{1}{H_0}
    \sum_{m,n}^\infty\fr{(-1)^{m+n}}{m!n!}\cd^{mn}_{\al,\be,\ga}\left(\fr{x(t)}{2}\right)^{2(m+n)+\al+\be+\ga},
 \ena
where the coefficients $\cc_{\al,\be,\ga}^{mn}$ and
$\cd_{\al,\be,\ga}^{mn}$ are defined by the following relations of
the Gamma functions
 \bega
    \cc_{\al,\be,\ga}^{mn}&=&\fr{\Ga(1+2m+\al+\be)}{(1-2m-\al-\be)
    \Ga(1+m+\al+\be)\Ga(1+m+\al)\Ga(1+m+\be)\Ga(1+n+\ga)},\\
     \cd_{\al,\be,\ga}^{mn}&=&\fr{\Ga(1+2m+\al+\be)}{(-2m-\al-\be)
    \Ga(1+m+\al+\be)\Ga(1+m+\al)\Ga(1+m+\be)\Ga(1+n+\ga)}.
 \ena
After some algebra, one arrives to the inhomogeneous solution for
the modes of the electromagnetic field $A^0$
 \beg
    X^{(0)}(t,k,\psi_0)=\fr{i}{\sin(\si\pi)}\sqrt\fr{\pi^3}{H_0}
    \left(\fr{2}{k}\right)^\fr{3}{2}
    \sum_{m,n}^\infty\fr{(-1)^{m+n}}{m!n!}\left(\fr{x(t)}{2}\right)^{2(m+n)}\sum_{s=1}^3{\cal
    E}^{mn}_{p_s}\left(\fr{x(t)}{2}\right)^{p_s},
 \en
where  $x(t)=k\,e^{-H_0 t}$. The sum over $s$ goes through three
different powers. The coefficients ${\cal E}^{mn}_{p_{s}}$ depend
on the parameters $\mu,\si$ and sum on indices $m,n$ in the
following way
 \bega
    && p_1={3\over 2}+\mu,\,\, \ce^{mn}_{p_1}=\fr{e^{\mu\pi i}}{\sin(\mu\pi)}\left[\fr{({3\over 2}-\mu)}{2}
    \left(-\cc_{\mu,\si,-\si}^{mn}+\cc_{\mu,-\si,\si}^{mn}\right)-
    \cd_{\mu-1,-\si,\si}^{mn}+\cd_{\mu-1,\si,-\si}^{mn}\right],\\\label{p2}
    && p_2={3\over 2}-\mu,\,\, \ce^{mn}_{p_2}=\fr{1}{\sin(\mu\pi)}\fr{({3\over 2}-\mu)}{2}
    \left(\cc_{-\mu,\si,-\si}^{mn}-\cc_{-\mu,-\si,\si}^{mn}\right), \label{us2} \\
    && p_3={7\over 2}-\mu,\,\, \ce^{mn}_{p_3}=-\fr{1}{\sin(\mu\pi)}
    (\cd_{1-\mu,\si,-\si}^{mn}-\cd_{1-\mu,-\si,\si}^{mn}).
 \ena
The inhomogeneous solution, $X^{j}(k,t,\psi_0)$, of $A^j$, has
essentially three contribution terms. For simplicity, lets split
the sources in the following way:
 \beg
    {\cal F}_1^{j}=2H_0^2(i\,k^j)\,W^{(0)}(k,t,\psi_0),\ \ \ \ {\cal
    F}_2^{j}=2(i\,k^j)\,H_0^3\phi(k,t,\psi_0),
    \ \ \ \ {\cal F}_3^{j}=2H_0^2(i\,k^j)\,X^{(0)}(k,t,\psi_0).
 \en
Hence, the final solution of (\ref{eu1}), written as
$X^{j}=X^{j}_1+X^{j}_2+X^{j}_3$, after using $k^j=k\,e^j$ ($e^j$
being an unitary vector), is given by the expressions
 \bega
    X_{1}^{j}(k,t,\psi_0)&=&-\fr{e^j}{\sin^2(\si\pi)}\sqrt\fr{\pi^3}{H_0}
    \left(\fr{2}{k}\right)^\fr{3}{2}\sum_{n}^\infty\fr{(-1)^{n}}{n!}
    \nonumber \\
    & \times &
    \left\{\fr{H_0t}{\Ga(1-\si)\Ga(1+\si)}\left[\fr{(x/2)^{\fr{5}{2}-\si}}{\Ga(1+n-\si)}
    +\fr{(x/2)^{\fr{5}{2}+\si}}{\Ga(1+n+\si)}\right]\right. \nonumber \\
    &+ &\sum_{m=1}^\infty\fr{(-1)^{m}}{m!}
    \left[\cd_{-\si,\si,-\si}^{mn}\left(\fr{x}{2}\right)^{\fr{5}{2}-\si+2m}+e^{\si\pi
    i}\cd_{\si,-\si,\si}^{mn}\left(\fr{x}{2}\right)^{\fr{5}{2}+\si+2m}\right] \nonumber \\
    &-&\left.\sum_{m=0}^\infty\fr{(-1)^{m}}{m!}\left[e^{\si\pi
    i}\cd_{\si,\si,-\si}^{mn}\left(\fr{x}{2}\right)^{\fr{5}{2}+\si+2m}+
    \cd_{-\si,-\si,\si}^{mn}\left(\fr{x}{2}\right)^{\fr{5}{2}-\si+2m}\right]
   \right\}\,\left(\fr{x}{2}\right)^{2n},\\
    X_{2}^{j}(k,t,\psi_0)& =& -\fr{e^j}{\sin(\si\pi)}\sqrt\fr{\pi^3}{H_0}
    \left(\fr{2}{k}\right)^\fr{1}{2}\sum_{m,n}^\infty\fr{(-1)^{m+n}}{m!n!}\left(\fr{x(t)}{2}\right)^{2(m+n)}
    \sum_{s=1}^2 2\ce^{mn}_{r_s} \left(\fr{x(t)}{2}\right)^{r_s},
 \ena
 with
 \bega
    r_1&=&3/2+\mu,\ \ \ \ \ \ \ \ \ \ \ce^{mn}_{r_1}=-\fr{e^{\mu\pi i}}{\sin(\mu\pi)}
    \left(\cc_{\mu,\si,-\si}^{mn}-\cc_{\mu,-\si,\si}\right),\\
    r_2&=&3/2-\mu,\ \ \ \ \ \ \ \ \ \ \ce^{mn}_{r_2}=\fr{1}{\sin(\mu\pi)}
    \left(\cc_{-\mu,\si,-\si}^{mn}-\cc^{mn}_{-\mu,-\si,\si}\right).
 \ena
Finally, the contribution of the inhomogeneous source is \bega
    X_{3}^{j}(k,t,\psi_0)=&&-\fr{e^j}{\sin(\si\pi)}\sqrt\fr{\pi^3}{H_0}
    \left(\fr{2}{k}\right)^\fr{1}{2}\fr{4\pi}{\sin(\si\pi)}\nonumber \\
&&\times\sum_{l,m,n,h}^\infty\fr{(-1)^{l+m+n+h}}{l!\,m!\,n!\,h!\,\Ga(1+l+\si)\Ga(1+h-\si)}
\left(\fr{x(t)}{2}\right)^{2(l+m+n+h)}\nonumber\\
&& \times \sum_{s=1}^3 \ce^{mn}_{p_s}\fr{\si\,
}{\left[p_s+2(l+m+n)-\fr{5}{2}\right]^2-\si^2}\left(\fr{x(t)}{2}\right)^{p_s}.
 \ena

\section{Calculation of the spectrum for the electric field fluctuations}\label{apb}

It is important to notice that $\dot X^j_1$ has appreciable
 differences with the other terms in (\ref{fou elec}); the one has a preceding factor
 $k^{-3/2}$, while the others ($iK^jX^0$, $\dot X^j_2$ and $\dot X^j_3$) are proportional to $k^{-1/2}$.
Furthermore,  $\dot X^j_1$ has terms with the linear factor $H_0t$
while the others doesn't.

 Taking into account the last observation we arrange the power spectrum as follows
 \beg\label{power1}
    {\cal P}_k(t)=k^2\sum_{q=0}^\infty a_q
    (ke^{-H_0t})^{\be_0+q}+k\sum_{q=0}^\infty
    b_q(t)(ke^{-H_0t})^{\ga_0+q}+\sum_{q=0}^\infty
    c_q(t)(ke^{-H_0t})^{\rho_0+q}.
 \en
The coefficients $a_q$ are all calculable from quadratics and
cross products of: $iK^jX^{0}$, $\dot X^j_2$ and $\dot X^j_3$. The
coefficients $b_q(t)=b^{(1)}_q+b_q^{(2)}t$ are linear in time and
come from products of $\dot X^j_1$ with the others. Finally, the
coefficients $c_q(t)=c^{(1)}_q+c^{(2)}_q t+c^{(3)}_q t^2$ are
quadratic in time and are found from $\sum_j|\dot X^j_1|^2$.
 The lowest powers from which each term in the series begin are:
 $\be_0=3-2\mu$, $\ga_0=-2+\mu+\si$ and $\rho_0=-4+2\si$.

 Since the terms that grow stronger are those that involve $a_q$,
 we shall restrict our study just to these. We shall try to obtain the power spectrum in a power-law form
 \beg\label{power2}
    {\cal P}_k(t)=C(t)k^{n_k-1}.
 \en
This will automatically lead us to a scale dependent spectral
index $n_k$, that it is found to be
 \beg
    n_k=3-\be_0+\fr{\ln\left[1+\sum_{i=1}^\infty\fr{a_i}{a_0}(ke^{-H_0t})^{\be_0+i}\right]}{\ln(k)},
 \en
where $C(t)=a_0e^{-\be_0H_0t}$ depends on the first coefficient
and the first power. We may write
 \bega
    n_k&=&n_0+\de n_k,\\
    n_0&=&3-\be_0 = 2\mu \simeq 3,\\
    \de
    n_k&=&\fr{\ln\left[1+\sum_{i=1}^\infty\fr{a_i}{a_0}(ke^{-H_0t})^{\be_0+i}\right]}{\ln(k)}.
 \ena
Since the values of $k$ are related to super Hubble wavelengths :
$0<k<\theta\si e^{H_0t}$ (and assuming that $\theta\si \ll 1$), we
see that $0<ke^{-H_0t} \ll 1$ and therefore it is pertinent a
perturbative analysis in powers of $\theta\si$. In this case the
dominant spectral index comes from $n_0=2\mu$, and $\de n_k$ are
perturbative corrections. The integration of any of the power
spectrums (\ref{power1}) or (\ref{power2}) provide us the
amplitude for electric fields
 \beg
    \int_0^{\theta k_H} \fr{dk}{k}{\cal P}_k(t)\simeq e^{2H_0t}\left[\fr{a_0(\theta\si)^{2+\be_0}}{2+\be_0}+\fr{a_1(\theta\si)^{3+\be_0}}{3+\be_0}+
                   \fr{a_2(\theta\si)^{4+\be_0}}{4+\be_0}+...\right].
                   \label{b7}
 \en
If we only stay with $n_0$, this would mean we are cutting the
previous expression just to the first term and only the
coefficient $a_0$ will appear. But if we keep to first order
corrections in $\de n_k$, we can see that the factor $a_1/a_0$
appears in the correction. In general we shall obtain that to
$N$th-order correction, the first $N$ coefficients will appear to
each order respectively.

In what follows we shall fix the spectral indices of the inflaton
as $\mu=3/2+\eps$, where $\eps=-m^2/2$ and $m^2$ is associated to
the measured spectral scalar index $n_s\simeq 0.96$
\cite{Spergel}: $m^2 \simeq -0.04$. The spectral index of the
vector fields is fixed so as to give a nearly scale invariant
spectrum of the magnetic fields $\si=5/2+\eta$, with
$\eta=-\nu^2/5$. For a similar spectrum to whole of the inflaton
field it is expected to $\nu^2$ to be negative, but $|\nu^2|\ll
1$.

Studying just the terms that grow faster, as $e^{2H_0t}$,  and
considering sufficiently large scales $\theta\ll1$, we shall cut
the power series to the first two terms. For the previous values
of $\mu$ and $\si$ there is no $a_1$,  since the power series
begins after the $a_0$ term, in $a_2$.

Since $(3/2-\mu)\ce^{mn}_{r_2}=2\ce^{mn}_{p_2}$, to obtain $a_0$
only we need to find $\ce^{00}_{p_2}$ in (\ref{us2})
 \beg
    \ce^{00}_{p_2}=\fr{\eps}{10\pi^{3/2}}\left[\fr{1}{\eps-\eta}-\fr{1}{5}\right],
 \en
and then
 \beg\label{apb9}
    a_0=\fr{H_0^4}{2\pi^2
    25}\eps^2\left[\fr{1}{\eps-\eta}-\fr{1}{5}\right]^2\left[1+\fr{2/5}{\eps-\eta}\right]^2.
 \en
Since both, $\eps$ and $\eta$ are respectively small departures
from $\mu$ and $\si$, then we obtain that $(\eps-\eta)^{-1}\gg1$.
We notice that $2/5(\eps-\eta)^{-1}$ comes from the solution
$X^j_3$, that considers only the contribution from the
inhomogeneous solution of $X^0$, coupled to the effective
inflaton. This means that here the most relevant solution is
$X^j_3$, and only considering this solution one obtains
 \beg
    a_0=\fr{2H_0^4}{\pi^25^4}\fr{\eps^2}{(\eps-\eta)^4},
 \en
with a spectral index $n_0=3$ and $\de n_k=0$.
\end{appendix}

\FIGURE[pos]{\epsfig{file=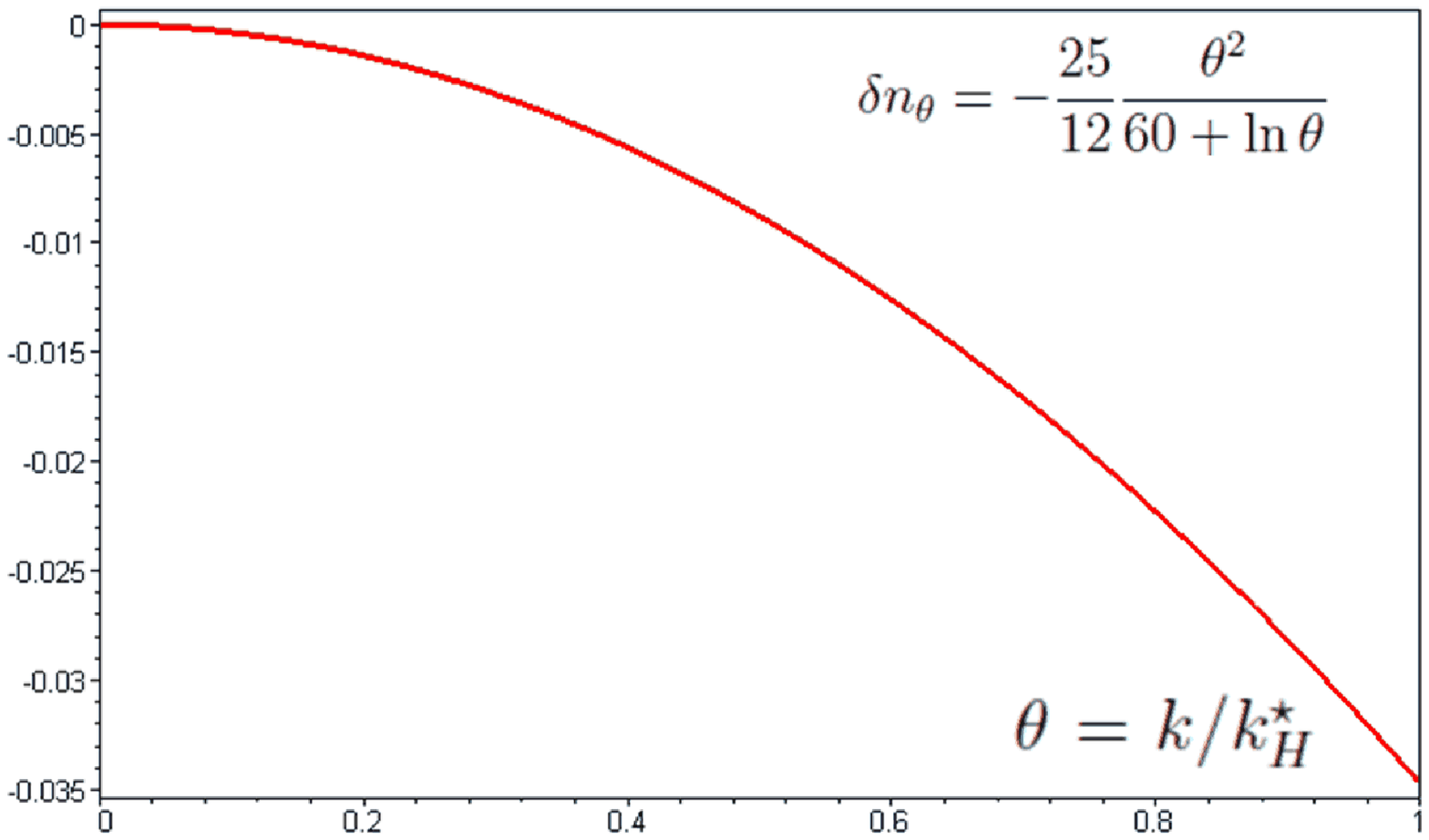,width=12cm,height=20cm}\label{f1}\caption{The
figure shows $\delta n_k$ as a function of $\theta=k/k_H$. Notice
that when the horizon entry (i.e., after inflation when
$\theta=\theta_*=k_*/k_{H_*}=1$), the value of $\delta n_k$ is
close to $\delta n_{k_*}\simeq -0.035$. In all our calculations we
have taken into account the value $k_{H_*} = \sigma e^{60}$ as the
wavenumber related to the horizon wavelength when, after
inflation, the horizon entry.}}
\end{document}